\documentclass[aps,pra,twocolumn,amsmath,nofootinbib,amssymb,superscriptaddress]{revtex4}

\newcommand{\bra}[1]{\langle#1|}
\newcommand{\ket}[1]{|#1\rangle}

\usepackage[pdftex]{graphicx}
\usepackage{mathrsfs}

\begin{document}

\bibliographystyle{apsrev}

\title{Error tolerance of the {\sc BosonSampling} model for linear optics quantum computing}

\author{Peter P. Rohde}
\email[]{dr.rohde@gmail.com}
\homepage{http://www.peterrohde.org}
\affiliation{Centre for Quantum Computation and Communication Technology, University of Queensland, Australia}
\affiliation{University of Paderborn, Applied Physics, 33098 Paderborn, Germany}
\affiliation{Centre for Engineered Quantum Systems, Department of Physics and Astronomy, Macquarie University, Sydney NSW 2113, Australia}

\author{Timothy C. Ralph}
\email[]{ralph@physics.uq.edu.au}
\affiliation{Centre for Quantum Computation and Communication Technology, University of Queensland, Australia}

\date{\today}

\frenchspacing

\begin{abstract}
Linear optics quantum computing (LOQC) is a promising approach to implementing scalable quantum computation (QC). However, this approach has very demanding physical resource requirements. Recently, Aaronson \& Arkhipov showed that a simplified model, which avoids the requirement for fast feed-forward and post-selection, while likely not capable of solving \textbf{BQP}-complete problems efficiently, can solve an interesting sampling problem, believed to be classically hard. Loss and mode-mismatch are the dominant sources of error in such systems. We provide evidence that even lossy systems, or systems with mode-mismatch, are likely to be classically hard to simulate. This is of practical interest to experimentalists wishing to demonstrate such systems, since it suggests that even with errors in their implementation, they are likely implementing an algorithm which is classically hard to simulate. Our results also equivalently apply to the multi-walker quantum walk model.
\end{abstract}

\maketitle

\emph{Introduction} --- Quantum computing (QC) \cite{bib:NielsenChuang00} offers exponential increases in computing power for certain key problems, however the building of a full-scale universal quantum computer remains well beyond current technological abilities. Nevertheless, single purpose machines capable of solving particular problems, may become possible on a shorter timescale. For example, the problem of building a quantum emulator - a well controlled quantum system whose dynamics approximate those of a classically intractable physical quantum system of interest - may be solvable in the medium term.

Linear optics QC (LOQC) \cite{bib:KLM01,bib:KokLovett11} has emerged as a leading candidate for the implementation of QC. In principle, universal QC operations can be implemented in several different ways using linear optics, photon production and counting, quantum memory and fast feedforward \cite{bib:Kok07}. Recently Aaronson \& Arkhipov (AA) \cite{bib:AaronsonArkhipov10} studied an alternate, but more straightforward approach to implementing LOQC. Here only photon preparation, passive linear optics elements and photo-detection are required. No fast feed-forward or quantum memory is necessary. AA showed that while this model probably cannot solve \textbf{BQP}-complete problems (i.e. it is likely not universal for QC), it can solve a sampling problem closely related to the \textbf{\#P}-complete matrix permanent problem, believed to be classically hard. That is, a passive linear optical network, fed with single photon states, can emulate a classically hard Boson-sampling problem.

An unanswered problem with quantum emulation is the effect of errors. In a universal QC, errors can be corrected via the implementation of fault tolerant codes. However, typically the restricted set of operations available to an emulator may be insufficient for fault tolerant error correction. Thus a pressing question for particular emulator proposals is whether they can still simulate classically hard dynamics in the presence of realistic levels of noise.

In this paper we address this problem with respect to the AA scheme and its tolerance to loss, i.e. detector, source and component inefficiency. This is the major error source for optical quantum processing. We present strong evidence that the AA scheme can still simulate classically hard problems in the presence of realistic inefficiencies. The trade-off compared to an error-free system is an increase in circuit size and photon number.

\emph{The Model} --- The AA model solves the so-called {\sc BosonSampling} problem by sampling the photon number configuration statistics at the output of a linear optics circuit with a multi-photon input state. We begin by preparing some number of modes $N$, where some configuration of $n<N$ modes are initialised with the single photon state $\ket{1}$, and the remaining \mbox{$N-n$} modes are in the vacuum state $\ket{0}$. Thus the input state is of the form \mbox{$\ket{\psi_\mathrm{in}} = \ket{1_1,\dots,1_n,0_{n+1},\dots,0_N}$},
where \mbox{$N=O(n^2)$}. The input state then passes through a passive linear optics network consisting only of beamsplitters and phase shifters, implementing a unitary map $U$ on the photon creation operators, \mbox{$a_i^\dag \to \sum_j U_{ij} a_j^\dag$}, where $a_i^\dag$ is the creation operator acting on mode $i$. Importantly, Reck \emph{et al.} \cite{bib:Reck94} showed that an arbitrary $U$ can always be decomposed into a polynomial number of optical elements. Thus any $U$ of this form can always be efficiently experimentally constructed.

\begin{figure}
\includegraphics[width=0.5\columnwidth]{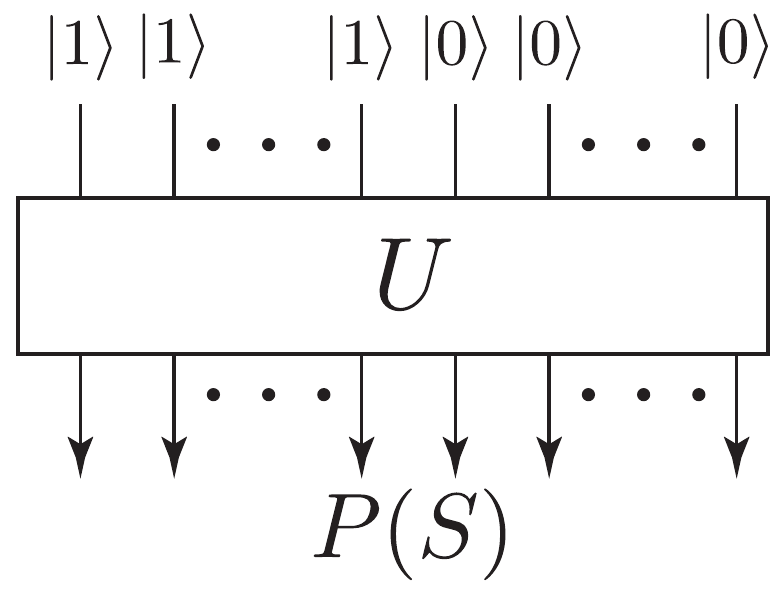}
\caption{The model for the {\sc BosonSampling} problem. We begin with $n$ photons, one in each of some number of modes, and none in the remaining modes. After the linear optics network, $U$, we measure some configuration of photons $S$.} \label{fig:model}
\end{figure}

In the occupation number representation the output state is of the form
\begin{equation} 
\ket{\psi_\mathrm{out}} = \sum_S \gamma_S \ket{n_1^{(S)},n_2^{(S)}\dots,n_N^{(S)}} 
\end{equation}
where $S$ are the different configurations of photons across the $N$ modes in the output state and $\gamma_S$ are the corresponding amplitudes. The model is illustrated in Fig. \ref{fig:model}. By detecting photon number at the output the Boson distribution for some particular input configuration and unitary scatterer, $U$, can be sampled. AA propose that if $U$ is picked randomly then it is not possible for a classical computer to efficiently simulate the sampling of the distribution.

We can formulate the problem in terms of a competition between two participants, Alice and Bob. Alice possesses a quantum emulator whilst Bob possesses a classical simulator. The adjudicator, Victor, draws a random unitary, $U$, from a hat and Alice and Bob race to sample its distribution to some predetermined accuracy in the shortest possible time. According to AA's thesis, for sufficiently large systems, Alice will consistently win the race.

\emph{Why Boson sampling is hard} --- AA demonstrate that should {\sc BosonSampling} be classically easy to simulate, this would have very surprising consequences in computational complexity theory, and therefore suggest that it is most probable that {\sc BosonSampling} is not classically easy to simulate \cite{bib:AaronsonArkhipov10}. From the complexity-class point of view an intuitive reason suggesting that {\sc BosonSampling} is classically hard is that each $\gamma_S$ is a function of a matrix permanent, \mbox{$P(S) \propto \mathrm{Per}(A_S)$}. The matrix permanent problem resides in the class \textbf{\#P}-complete, strongly believed to be classically hard. Importantly though, {\sc BosonSampling} does not allow us to \emph{calculate} matrix permanents efficiently, since doing so would require an exponential number of measurements. Rather, it samples across many different matrix permanents.

More pragmatically, we can look at particular classical simulation approaches and see how they fail. Considering the size of the Hilbert space it can easily be seen that, with indistinguishable photons, the number of parameters $|S| = \binom{N+n-1}{n}$ grows exponentially with the number of photons and modes in the system. The emphasis on indistinguishable photons is important. If the photons are distinguishable then the output can be shown to factor into:
\begin{equation} 
\ket{\psi_\mathrm{out}} =\prod_i \sum_{S'}\gamma^i_{S'} \ket{n_{1i}^{(S')},n_{2i}^{(S')}\dots,n_{Ni}^{(S')}} 
\end{equation}
where now $\gamma^i_{S'}$ are the amplitudes for the different outputs a single input photon can reach, with the product over all $n$ input photons. For each photon there are only $S'=N$ output configurations so the total number of independent parameters is $n \times N$.  As this only grows polynomially for distinguishable input photons, such a situation is classically simulatable. We will return briefly to this issue later when we consider mode-matching. AA suggest that a 400 mode interferometer fed with 20 single photons lies around the boundary of the simulation powers of current classical computers. For this case we find $|S| \approx 7 \times 10^{33}$.

The problem might also be simulated in the quadrature field basis. It is known that such a simulation will be efficient if the input state is Gaussian \cite{bib:Bartlett02,bib:Bartlett02b} and the photon counting is not adaptive\footnote{An adaptive measurement scheme is one in which the type of measurement made on the ($i+1$)th mode is a function of the measurement result obtained from the $i$th mode. In a non-adaptive measurement scheme, such as in the AA proposal, all the measurement bases are predetermined.}. Single photon states are non-Gaussian but have a finite overlap with Gaussian states. This motivates the question: how close can a Gaussian simulation be to the {\sc BosonSampling} problem? In particular, suppose we replace each single photon state at the input with a Gaussian state and then simulate the unitary evolution of interest with this state. How similar can the two output states be? 

A measure of the similarity of  two density operators is the trace distance, $D=\frac{1}{2}\mathrm{tr}|\rho - \sigma|$. The trace distance is a metric. It equals zero when the states are identical and one when they are orthogonal. 
It is monotonically related to the lower bound of the Shannon mutual distinguishability between the two density operators \cite{bib:Fuchs97}. Thus, given the symmetry of the physical situation we consider\footnote{The density operators are symmetric in the sense that we assume every single photon state input in $\rho$ is replaced by the optimal Gaussian approximation in $\sigma$. Thus the distortion of $\sigma$ away from $\rho$ is symmetric over all the modes.}, the trace distance can bound the amount of information that can be learned about density operator $\rho$ via a simulation which produces density operator $\sigma$. 
Furthermore, the trace distance is invariant under unitary transformations, thus by computing the minimum trace distance between the input single photon distribution and a Gaussian input state we can evaluate the similarity between the quantum emulation and a classical simulation of the output for arbitrary unitaries. 

The trace distance between a single photon state and single Gaussian state is easily calculated and is minimized by maximizing the absolute value of the single photon amplitude of the Gaussian. We obtain $D=0.522$ for the squeezed state with amplitude $\beta=\sqrt{2}$ and squeezing of $V=1/3$ (where the quantum noise limit is $V=1$) \cite{bib:yue76}. 
For this simple case of pure states the trace distance between $n$ copies of the single photon state and $n$ copies of the minimizing Gaussian state is $D_n = 1-(1-D)^n$, which asymptotes to unity exponentially in the number of copies. Thus we find that a Gaussian simulation of the problem will rapidly become useless as the size of the system grows. Considering again the case of 20 input photons, and using the above distance minimising classical state, we find $D_n = 1.0 - 3.86 \times 10^{-7}$ - clearly very close to one, implying that the Gaussian simulation would give virtually no information about the {\sc BosonSampling} problem.   

\emph{Effect of Loss} --- Having gained some physical intuition into why the {\sc BosonSampling} problem is hard we now wish to investigate the effect of errors, in particular photon loss, on the operation of the quantum emulation. In Appendix A we present an argument based on truncation of the photon number basis in a classical simulation to show the problem becomes ``easier" as loss increases. Nevertheless, if only small amounts of loss are present then it may still be reasonable to post-select successful events in a probabilistic way. Provided the overhead is not too large the AA system can still be said to be solving a hard problem efficiently. 

Consider the case of the 20 input photons to the 400 mode interferometer. Let us assume the total efficiency for a single photon to be produced, transmitted through the interferometer and to be successfully detected is $\eta$. Let us further assume that this efficiency is the same for all photons. Under such conditions the detection and transmission efficiencies can be commuted back to the source and lumped together as a single source efficiency \cite{bib:Lund08,bib:Varnava07}. The chance that all 20 photons will be detected at the output is then $P=\eta^{20}$. Provided this number is not too low then statistics can still be obtained for the sampling problem by running the system many times and only keeping those cases where 20 photons are actually detected. Suppose we conservatively require $P=0.5$, meaning we need to run the system about twice as many times as the no loss case to obtain good statistics. We conclude that a bulk efficiency of about $\eta \geqslant 0.96$ is needed. Lower efficiencies are allowed if we allow for more repetitions of the experiment, however if Alice postselects too much she will lose her advantage over Bob's classical processor. The figure of $0.96$ is good compared to thresholds for universal quantum computing where efficiencies higher than $0.99$ are typical, and clearly illustrates the advantage of single purpose machines. However, this still represents a major challenge given current technology. In the following we consider a different direction, more promising for the short term.

\emph{Is lossy Boson sampling a hard problem?} ---  We now address the question `can lossy AA systems be efficiently classically simulated?', and present evidence that this is not the case in situations of near term experimental interest. Although we have seen that if the efficiency is too low then the probability of post-selecting the samples of a lossless scattering problem are very low, it may still be true, for moderate levels of loss, that the lossy {\sc BosonSampling} problem {\it itself} is hard to simulate on a classical computer - and thus of interest for quantum emulation. We base our evaluation on whether the problem remains hard for the two simulation approaches discussed earlier. We observe that if the average photon number of the input state is equal to, or greater than that for the lossless case, with an equal total number of modes, then the number of parameters, $|S|$, required for a direct number state simulation will be at least as large as for the lossless case. On the other hand we recognize that a lossy single photon state is more Gaussian than a pure one, and so we could expect a Gaussian simulation of lossy {\sc BosonSampling} to be a better approximation than it would be for lossless {\sc BosonSampling}. However, we also know from the previous section, that for small amounts of loss the AA system remains hard to simulate due to the ability to post-select. By calculating the trace distance between the output state of an AA system with a small amount of loss  and the corresponding closest Gaussian state we can obtain a benchmark condition for deciding if a Gaussian simulation will be useful in more general situations. In summary we seek lossy AA systems that fulfil the following criteria:

(i) $\bar n_L \geqslant n$, where $\bar n_L$ ($n$) is the average photon number for the lossy (lossless) case and the total number of modes is of $N=O(n^2)$; and

(ii) $D_L \geqslant D$, where $D_L$ is the trace distance between the output of a lossy AA system and the closest equivalent Gaussian state, whilst $D$ is similar but for an AA system for which the loss is sufficiently small that the post-selection probability $P \geqslant 0.5$. We particularly consider the case of $n=20$ for our calculations.

A lossy single photon state can be represented as a density operator of the form \mbox{$\rho_\mathrm{lossy} = (1-\eta) \ket{0}\bra{0} + \eta \ket{1}\bra{1}$}. In general the closest Gaussian state will be a phase diffused, displaced squeezed state. The expression for the trace distance between $n$ lossy single photon inputs and $n$ phase diffused, displaced squeezed state becomes:
\begin{eqnarray}
D_n &=& {{1}\over{2}} \sum_k \binom{n}{k} |(1-\eta)^{n-k} \eta^k - {C_0}^{n-k} {C_1}^k | \nonumber\\
&+& {{1}\over{2}}(1-(C_0+C_1)^n)
\end{eqnarray} 
where
\begin{equation}
C_i = \frac{1}{i! \mu} \left(\frac{\nu}{2\mu}\right)^i \left|H_i\left(\frac{\beta}{\sqrt{2\mu\nu}}\right)\right|^2 e^{-\beta^2(1-\nu/\mu)},
\end{equation}
is the $ith$ photon number probability of a phase diffused, displaced squeezed state with displacement $\beta$ and squeezing $V = (\mu-\nu)^2$ (with $\mu^2-\nu^2 =1$) , and $H_i$ are Hermite polynomials. The trace distance can be minimized as a function of $\beta$ and $V$ to give $D_n^\mathrm{min}$. In particular we find that for $n=20$ and $\eta=0.96$ that $D_n^\mathrm{min} = 0.9987$ for $\beta=1.217$ and $V=0.3801$.

\begin{figure}[!htb]
\includegraphics[width=0.9\columnwidth]{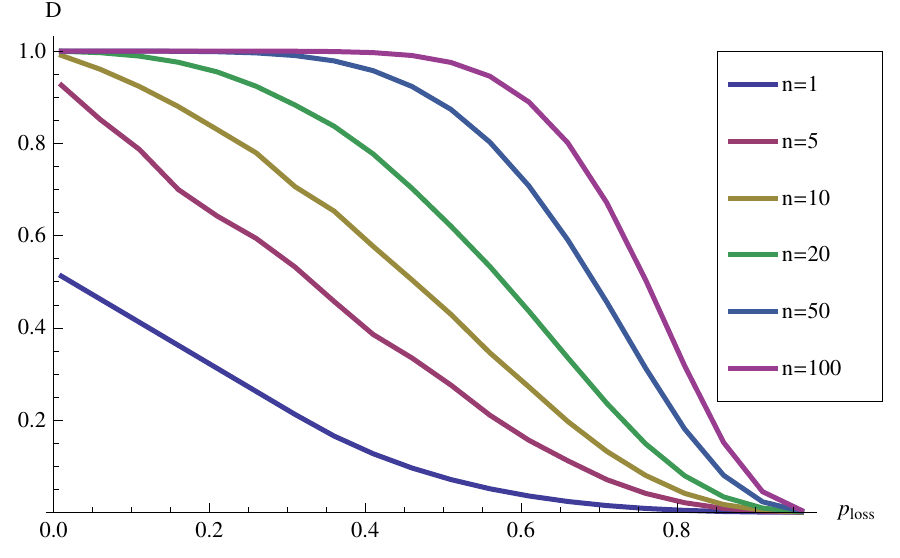}
\includegraphics[width=0.95\columnwidth]{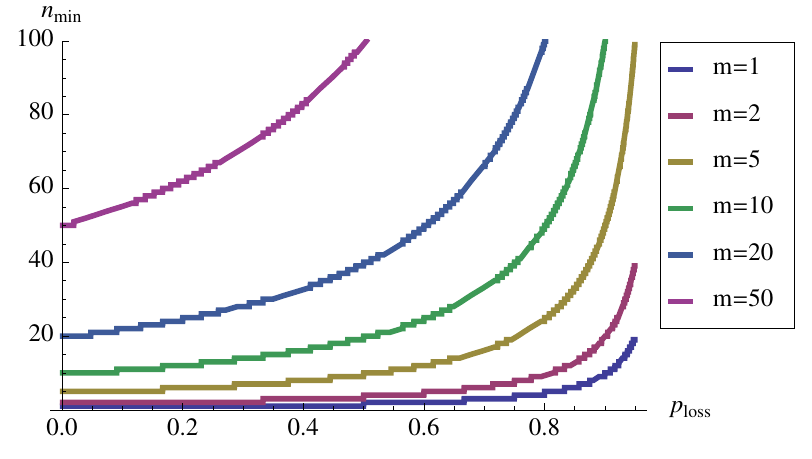}
\includegraphics[width=0.8\columnwidth]{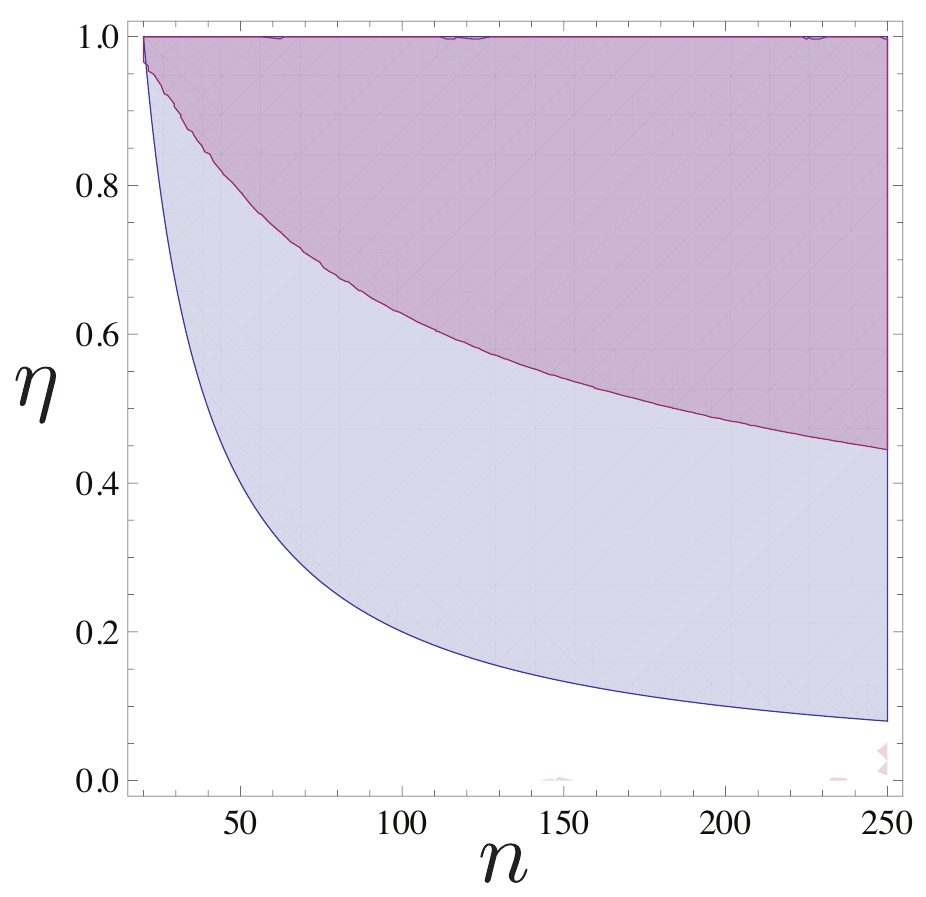}
\caption{Colour online. (top) Minimum trace distance between the density operators of the actual lossy system compared to a Gaussian simulation using an optimised squeezed state for different input photon numbers $n$. (middle) Physical resource requirements, as measured by the required number of photons, such that the average number of photons in the system is $m$. (bottom) The region satisfying our conditions (i), in blue, and (ii), in red. The overlapping region is presumed to be hard to classically simulate.
}. \label{fig:lossy_simulation}
\end{figure}

In Fig. \ref{fig:lossy_simulation}(top) we plot $D_n^\mathrm{min}$ against the loss rate $\eta$ for various input photon numbers, $n$. Two trends are clear. For a fixed input photon number, increasing loss leads to decreasing minimum trace distance, indicating that the Gaussian simulation is becoming a better approximation to the quantum emulation. On the other hand, for fixed loss, increasing the input photon number increases the minimum trace distance, indicating that the Gaussian simulation is becoming a worse approximation to the quantum emulation. This suggests that the ``hardness" of the simulation can be maintained in the presence of increasing loss by simultaneously increasing the input photon number.

In Fig. \ref{fig:lossy_simulation}(middle) we plot the number of photons that must be input into the system such that the average photon number after loss, $\bar n_L = m$, and hence the number of configuration amplitudes that would need to be calculated via a number state simulation remains approximately constant. Again we find that the ``hardness" of the simulation can be maintained if we compensate for loss by injecting additional photons. 

In Fig. \ref{fig:lossy_simulation}(bottom) we plot the regions satisfying our conditions (i) and (ii), i.e. the regions where the ``hardness" of the simulation, with respect to photon number and Gaussian approaches, is maintained at the same level as for $n=20$, $\eta=0.96$. We find that for large photon numbers the requirements for a hard calculation are rather lenient. In particular, loss levels can exceed 50\% for $n\gtrsim 200$.

\emph{Mode-mismatch} --- In addition to loss, mode-mismatch is another dominant source of error in AA type systems. Using the techniques of Ref. \cite{bib:RohdeMauererSilberhorn07}, we can decompose the spectral distribution function of the input photons into overlapping and non-overlapping components, governed by a distinguishability parameter $\lambda$. As $\lambda$ grows the number of independent configurations drops as the number of distinguishable photons increases.

The standard technique to overcoming mode-mismatch is using narrowband filtering. This technique is widely used in virtually all LOQC experiments. The idea here is that if we project onto a narrow window in say frequency-space, then within that window distinguishable photons will appear indistinguishable. The introduction of filtering is equivalent to loss, since we are discarding a portion of the wave-function. Specifically, after filtering onto some small window $w$ around $\omega_0$, the probability of detecting the photon is \mbox{$p(\omega_0) = \int_{\omega_0}^{\omega_0 + w} |\psi(\omega)|^2 d\omega$}, where $\psi(\omega)$ is the spectral distribution function of the photons. This is equivalent to a lossy channel with loss probability \mbox{$1-p(\omega_0)$}. Thus, the problem of mode-mismatch is reduced to the problem of loss. Our results then indicate that some level of mode mismatch can be tolerated provided the total loss budget, including filtering, does not exceed the amounts allowed by Fig. \ref{fig:lossy_simulation}(bottom).

\emph{Relation to multi-walker quantum walks} --- Quantum walks (QWs) \cite{bib:ADZ,bib:AAKV,bib:Childs09,bib:Kempe08} have recently obtained a lot of interest as an approach to quantum information processing. It was noted in Ref. \cite{bib:RohdeComment10} that to achieve exponential speedup in quantum algorithms, multiple walkers (photons) must be introduced into a QW system (assuming non-exponentially sized graphs). A formalism for multi-walker walks was introduced by Rohde \emph{et al.} \cite{bib:RohdeSchreiber10}. Indeed multi-walker systems have begun to be experimentally demonstrated \cite{bib:Peruzzo10,bib:Owens11}. The difference between general linear optics networks and the QW formalism is the presence of \emph{coins} in quantum walks. However, it can be shown \cite{bib:RohdeFedrizzi11} that a multi-walker quantum walk can be directly mapped to a beamsplitter network with quadratic resource overhead. A reverse mapping also always exists. Thus, QWs can be regarded as isomorphic to the formalism presented by AA, and the results we present here apply equally to the multi-walker QW formalism (in the absence of feed-forward). Furthermore, AA suggest that it is unlikely that general LO networks are universal for QC, implying the multi-walker QW formalism is also not universal.

\emph{Conclusion} --- We have considered the emulation of the {\sc BosonSampling} model with LOQC and investigated the effect of loss and mode-mismatch. We have presented evidence that lossy systems remain hard to simulate provided the loss is compensated through the injection of higher photon number input states. This observation should be of interest to quantum optics experimentalists, for whom experimental implementations of AA systems are in principle viable, but where loss is a ubiquitous problem. These results suggest that even with significant loss in their systems, they may be able to experimentally implement a quantum emulation in optics which could consistently beat the best classical simulation.

\begin{acknowledgments}
This research was conducted by the Australian Research Council Centre of Excellence for Quantum Computation and Communication Technology (Project number CE110001027).
\end{acknowledgments}

\bibliography{paper}

\section{Appendix}



\subsection{Simulation via truncation}

We consider an alternate approach to classically simulating a lossy AA system: via truncation of the classical simulation to a given number of photons and discarding higher photon number terms. Defining an error metric as the distance between the probability distributions for the ideal and truncated simulations, we find that
\begin{equation}
\epsilon = \sum_{i=m+1}^n \binom{n}{i}{p_\mathrm{loss}}^{n-i}(1-p_\mathrm{loss})^i \left[1-(1-1/N)^i\right],
\end{equation}
where $m$ is the number of photons the simulation is truncated to, and we have made the assumption that the network is approximately balanced (i.e. an incident photon has approximately the same probability of reaching any given output).

Fig. \ref{fig:lossy_simulation}(bottom) illustrates the error measure against the number of photons the simulation is truncated to. When there is no loss in the quantum system, the classical simulation is a very poor approximation unless we simulate the full $n$ photons, which is presumed to be exponentially hard via the results of AA. As the loss rate increases we can tolerate greater levels of truncation in the classical simulation, making it more tractable. 

As before, when the loss rate is very high, we are effectively simulating the vacuum state, which is computationally easy. However, for moderate loss rates, where there are still a large number of photons in the system, the system cannot be classically simulated via truncation unless the level of truncation is very small.
\begin{figure}[!htb]
\includegraphics[width=\columnwidth]{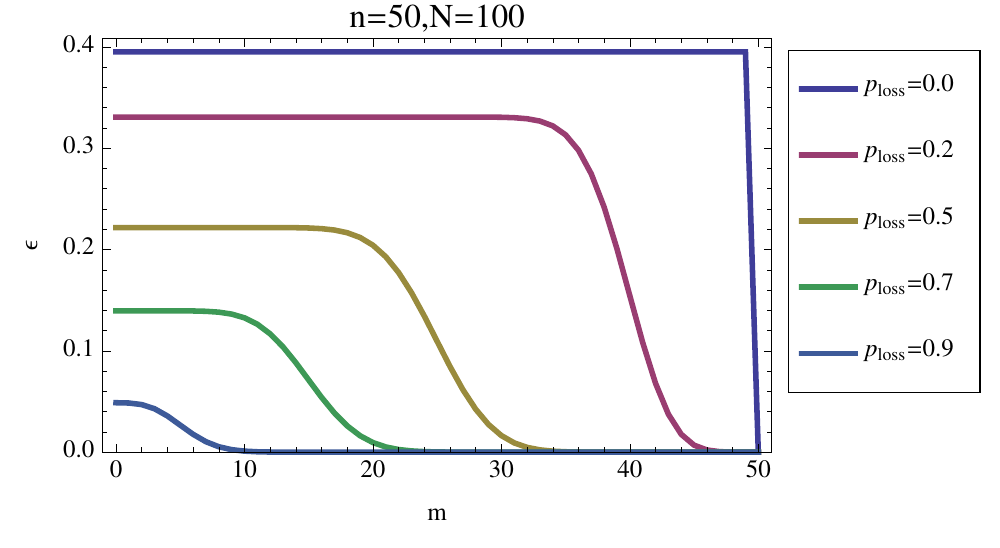}
\caption{Colour online. 
Simulating a lossy system via truncating a classical simulation to $m$ photons and discarding higher photon number terms.}. \label{fig:lossy_simulation_trunc}
\end{figure}

\end{document}